\documentclass[aps,pra,showpacs,amsmath,amssymb,twocolumn]{revtex4-1}

\usepackage{graphicx}
\usepackage{dcolumn}
\usepackage{bm}
\usepackage{enumerate}
\usepackage{amsmath}
\usepackage{amssymb}
\usepackage{hyperref}
\usepackage{color}
\usepackage{times}
\usepackage{physics}

\allowbreak

\begin{document}

\title{Open and closed spin chains as multiprocessor wires: optimal engineering and reachability}

\author{N. E. Palaiodimopoulos}
\email[]{nikpalaio@phys.uoa.gr}
\affiliation{Department of Physics, National and Kapodistrian University of Athens, GR-15784 Athens, Greece}

\author{I. Brouzos}
\affiliation{Department of Physics, National and Kapodistrian University of Athens, GR-15784 Athens, Greece}

\author{N. Georgoulea}
\affiliation{Department of Physics, National and Kapodistrian University of Athens, GR-15784 Athens, Greece}

\author{P. A. Kalozoumis}
\affiliation{Materials Science Department, School of Natural Sciences, University of Patras, GR-26504 Patras, Greece}
\affiliation{Hellenic American University, 436 Amherst st, Nashua, NH 0306 USA}

\author{F. K. Diakonos}
\affiliation{Department of Physics, National and Kapodistrian University of Athens, GR-15784 Athens, Greece}

\date{\today}

\begin{abstract} 

We consider the perfect transfer of a state between arbitrary nodes of one-dimensional spin-1/2 chain with optimally engineered couplings. Motivated by the fact that such a system could be used as a data bus for connecting multiple quantum processors, we derive two necessary and sufficient conditions that have to be met in order to perfectly transfer a state between any two nodes and we employ them to examine both open and closed geometries. Analytical calculations and numerical optimizations are performed for both cases in order to determine the reachability of certain target states and to provide optimal values for the couplings which ensure perfect fidelity. An important finding is that even-sized closed chains allow for perfect transfer between any pair of sites and therefore are a promising platform for the implementation of data bus protocols.

\end{abstract}

\maketitle

\section{Introduction}

The ability to faithfully transfer quantum states in a network of quantum processors is a crucial task that needs to be addressed, in order to construct an efficient platform for quantum computation ~\cite{divincenzo}. In the seminal paper of Bose ~\cite{bose1}, a one dimensional spin chain was proposed to act as a data bus for the reliable transfer of a quantum state over short distances. Two are the main advantages of the aforementioned scheme: (i) First there is no need for dynamical control since the system evolves freely to the desired state. (ii) Since the bus geometry and the quantum processor are build up by the same ingredients we avoid mapping the quantum state to photons, which so far seem to be the most promising candidates for long distance quantum communication. Both the mapping and the dynamical control are responsible for errors that arise during the transfer process. 

Bose's initial proposal considered a spin chain with uniform couplings between nearest neighbors. In this scenario, the maximum length of the chain that can support transfer with fidelity equal to unity, which is commonly referred as Perfect State Transfer (PST), is $N=3$. For longer chains information transfer still surpasses the classical limit but it is not perfect. However, further studies ~\cite{christandl1,christandl2} showed that when the fixed couplings between adjacent sites are engineered in a suitable manner then the quantum state can be perfectly transferred from one end to the other for chains of arbitrary length. The optimal profile for the couplings has also been obtained when studying the coherent transport of an electron in a chain of coupled quantum dots ~\cite{nikolopoulos1}. A recursive formula for constructing the optimal profile for the couplings has been introduced in ~\cite{wang}. Besides, it has been analytically demonstrated that a periodic profile of the couplings together with a constraint on the energy spectrum of the Hamiltonian are the sufficient and necessary conditions for PST between the end sites of the chain ~\cite{vinet1}. Moreover, it has been shown, that PST is in general possible between mirror symmetric sites of the spin chain ~\cite{albanese1}. Finally, experimental results supporting the efficiency of the protocol have been obtained by employing evanescently coupled waveguides ~\cite{bellec,perez,chapman}.    

In the current study we consider one-dimensional spin chains with nearest neighbor Heisenberg interactions, for open and closed geometries. Instead of considering the end nodes of the chain as initial and target sites, we aim to construct a data bus that can perfectly transfer a state from any initial to any target site assuming all are connected to a quantum processor. From further analytical calculations we perform, it is clear that the two conditions we have derived, can not be met simultaneously for every initial and target site that we pick. For open chains, we can firmly exclude certain cases of short distance transfer and for odd-sized chains cases where the state is transfered from a site with an even index to an odd one or vice versa. For cases where both conditions can be met we provide both an analytical scheme and numerical calculations to access the optimal profile for the couplings, which, in general, is not periodic. For closed chains, which are essentially one-dimensional circular ring geometries, we have opposite behavior for odd- and even-sized systems. For odd-sized chains ($N>3$) independently of the initial and target sites no PST is possible. On the contrary, we have observed that for even-sized chains there is always an optimal configuration that can support PST between arbitrary initial and target sites. Conclusively we highlight the advantage of an even-sized ring geometry for connecting multiple quantum processors to each other.

\section{A one dimensional data bus} \label{matrix}
\noindent 

We consider the one-dimensional XX Heisenberg spin-$1/2$ chain of length $N$, described by a Hamiltonian matrix follows:
\begin{equation} \label{eq:heis}
\mathcal{H}=\frac{1}{2} \sum_{i=1}^{N-1} J_{i} (S_{i}^{x} S_{i+1}^{x}+S_{i}^{y} S_{i+1}^{y}),
\end{equation}
where $J_{i}$'s correspond to the couplings betweeen adjacent sites and $S_{i}$'s to the spin operators on each lattice site. We have assumed that the $J_{i}$'s are real and positive, the interaction takes place only between nearest neighbors and that the magnetic field is absent. Without loss of generality one may also have a homogeneous magnetic field with equal strength on each lattice site. The Heisenberg Hamiltonian of Eq. (\ref{eq:heis}) expressed in the site basis can be represented as a block diagonal matrix, where each of the $N+1$ blocks corresponds to a fixed number of up-spins. The protocol we will consider starts with an initial state that has only one spin pointing up and our aim is to perfectly transfer this excitation to an arbitrary lattice site. Since the block corresponding to one spin up is not connected to the others, the time evolution operator, up to a phase, only affects this block. Thus, we can restrict ourselves to the one-excitation subspace where the Hamiltonian is represented by a real symmetric matrix that takes the following form
\begin{equation} \label{eq:matrix}
\mathcal{H}_{N}=
\begin{pmatrix}
0 & J_{1} & 0 & \dots & & J_{N}
\\ J_{1} & 0 & J_{2} &  &  & \vdots
\\ 0 & J_{2} & 0 & \ddots &  &
\\  \vdots &  & \ddots &\ddots   &  &
\\ & & & & & J_{N-1}
\\    &   &   &  &  & 
\\  J_{N} & \dots  &  &  & J_{N-1}  & 0 
\end{pmatrix}
.
\end{equation}
For open chains we must set $J_{N}=0$, so that the first and the last site get disconnected. It is worth noting here, that our results hold for every Hamiltonian that can be reduced to this particular matrix form. Many studies ~\cite{bose1,christandl1,vinet1} have initially considered different initial Hamiltonians for this protocol, but since only the one-excitation subspace evolves non-trivially, irrespectively of the initial Hamiltonian we always end up with the above matrix describing a single spin moving along the chain. Also, without loss of generality, the initial state may be a superposition between the one-excitation and the non-excited state where all spins are down. These two states are completely decoupled. Therefore, one transfers the "superposition" or an arbitrary qubit state, just by ensuring that the one-excitation state will be transferred perfectly.

\section{Perfect state transfer} \label{reach}
Two key notions are important to define in order to set the frame of our study: reachability and fidelity. The problem of reachability is to define sets of quantum states (initial and target) that are connected through time evolution. This means that we can transfer the initial state fully to the target one in a finite time, if we design a specific time-evolution operator -- which can be time-dependent or time-independent-- within the conditions and restrictions of the problem. Here we restrict our selves to time-independent Hamiltonians of the form of Eq. (\ref{eq:heis}), with finite $J$ values. Fidelity measures how faithfully we can transfer a quantum state, in a finite amount of time $\tau$, that is initially localized on one lattice site $\ket{m}$ to another $\ket{n}$, where $m,n=1, \dots, N$ denote the site basis vectors. 
\begin{equation} \label{eq:fidelity}
\mathcal{F}=\abs{\bra{n}e^{-i \tau \mathcal{H}_{N}}\ket{m}}^{2},
\end{equation}
 For $m=n$ we retrieve the probability of revival, where the system returns to its initial state. Revivals in this system are always possible due to the quantum recurrence theorem ~\cite{bocchieri}.

Fidelity is a function of the couplings $J_{i}$ and time $\tau$, $\mathcal{F}=f(J_{i}, \tau)$, where $i=1,..,N-1$ for open chains. The idea of tuning the couplings to achieve PST, can be seen as a process, where, in this $N$-dimensional ($N+1$ for closed chains) parametric space, we try to identify the set of values such that $\mathcal{F}=1$. A straightforward ab initio approach, followed here, is to use a numerical optimization search algorithm in order to find the parameter set that maximizes fidelity. For a chain of fixed length, when we want to transfer a quantum state that is initially localized on one lattice site to another, with $\mathcal{F}=1$, we quickly notice that no matter how extensively we search at the parametric space at some cases we are unable to do so. At this point, two are the main questions that need to be adressed. Firstly, which are the reachability criteria that have to be met for a PST to occur and secondly why these criteria cannot be met in several cases even thought the parametric space seems ``vast".

\subsection{Reachability criteria}

To answer the aforementioned questions we will derive the necessary and sufficient criteria for PST. In contrast to previous studies the criteria that will be presented here are general and do not only hold for transferring an excitation between mirror-symmetric lattice sites but also for arbitrary initial and target sites.

The Hamiltonian matrix (\ref{eq:matrix}) has a discrete symmetric non-degenerate spectrum and the eigenvalue equation writes as follows
\begin{equation}
\mathcal{H}_{N} v_{i}=E_{i}v_{i},
\label{eq:eigeneq}
\end{equation}
where $E_{i}$'s are the eigenenergies and $v_{i}$'s the N-component eigenvectors $v_{i}=(v_{i1},v_{i2},...,v_{iN})$, with $i=1,..,N$. Suppose that the system is initially prepared in state $\ket{m}$ and we want to transfer the excitation to a state $\ket{n}$ in a finite amount of time. Since the system evolves freely, the probability amplitude of finding the system on site $n$ after time $t$, in terms of the eigenvector components, is given by:
\begin{equation}
\label{eq:amp}
 \bra{n}e^{-i \tau \mathcal{H}_{N}}\ket{m}=\sum_{i=1}^{N} v_{im} v_{in} e^{-i \phi_{i}}
\end{equation}
where $\phi_{i}=tE_{i}$. In order for this sum to be equal to 1, it is required that $\abs{v_{im}}=\abs{v_{in}}$. This is the first criterium for reachability and its mathematical proof is given in Appendix \ref{proof}. Furthermore, the phases $\phi_{i}$ have to be such that they produce an overall plus or minus sign to the N-terms of the sum. The demand that PST occurs in a finite time results to a constraint on the energy spectrum. This emerges in the following manner. Without loss of generality, since $\phi_{i}$ is a product of $t$ and $E_{i}$ we can choose the retrieval time to be $t^{*}=\pi /2$ or $t^{*}=\pi$. Consequently, the energies are forced to take the following values
\begin{equation}
\label{eq:cases}
E_{i}=
\begin{cases}
n_{i}\\
2n_{i}+1
\end{cases}
, n_{i}=0,1,2,...
\end{equation} 
Thus, we get that the fraction of two eigenvalues always has to be a rational number. This constitutes the second reachability criterium. Henceforth, we will refer to the latter as the ``rationality" criterium.  

\subsection{Open chains}
To address the second question i.e., when the reachability criteria can be met, we will now explicitly demonstrate that for certain cases the two criteria for PST cannot be met simultaneously. To do so, we first focus our study on open chains. The eigenvalue condition [Eq. (\ref{eq:eigeneq})] is a linear system of $N$ equations, where, each eigenvector component is defined up to an arbitrary sign $s_{ij}$.
\begin{equation}
\label{eq:sysopen}
\begin{array}{lcr}
J_{1}s_{i2}\abs{v_{i2}}  = E_{i}s_{i1}\abs{v_{i1}} \\
J_{1}s_{i1}\abs{v_{i1}}+J_{2}s_{i3}\abs{v_{i3}}  =  E_{i}s_{i2}\abs{v_{i2}}\\
\vdots \\
J_{N-2}s_{iN-2}\abs{v_{iN-2}}+J_{N-1}s_{iN}\abs{v_{iN}}  =  E_{i}s_{iN-1}\abs{v_{iN-1}}\\
J_{N-1}s_{iN-1}\abs{v_{iN-1}}=E_{i}s_{iN}\abs{v_{iN}},
\end{array}
\end{equation}
To deduce whether PST between the first and the $m$-th site is possible we will exploit the first $m-1$ equations of the linear system Eq. (\ref{eq:sysopen}). By doing so, we can express $\abs{v_{jm}}$ as a function of $\abs{v_{j1}}$ as follows 
\begin{equation}
\label{eq:energy}
\abs{v_{im}}=\frac{E^{n-1}-E^{n-3}_{n \geq 3} \sum\limits_{j}^{n-2}J_{j}^{2}+E_{n\geq 5}^{n-5} \sum\limits_{j \neq k}^{n-2} J_{j}^{2}J_{k}^{2}+...}{s_{1m}\prod\limits_{j}^{n-1}J_{j}}\abs{v_{i1}},
\end{equation}
where $s_{1m}$ is the relative sign between the first and the $m$-th component of $v_{j}$.Then by employing the first reachability criterium we can set $\abs{v_{im}}/\abs{v_{i1}}= 1$ and we end up with two energy polynomials, corresponding to the plus or minus sign of the product in the denominator. The number of the real roots of the two energy polynomials added together, has to be greater or equal to the total number of the system's eigenvalues otherwise PST can not be achieved.

Based on this counting argument, it is straightforward to deduce that PST from the first site to any target site $n$, when $n \leq N/2$ for even-sized and $n \leq (N+1)/2$ for odd-sized chains, is forbidden. Additionally, since an open chain is mirror symmetric around the axis that passes from its center, two mirror symmetric transfer processes have the same properties. For example, when we consider the transfer from the first to the third site of a 6-site chain, based on the above, we have two second degree polynomials that can give four roots. Thus, PST cannot be made possible, since we ought to have at least six roots. By invoking the mirror symmetry of the chain, the same holds for the transfer between the fourth and the sixth site of the chain.   

Moreover, specifically for odd-sized chains,  because the spectrum of the Hamiltonian is symmetric there will always be a zero energy eigenvalue. If we want to examine whether PST can occur from the $n$-th to the $m$-th site we can use Eq. (\ref{eq:energy}) for  $\abs{v_{in}}$ and $\abs{v_{im}}$. Using these two relations we can expunge  $\abs{v_{i1}}$ and express $\abs{v_{im}}$ as a function of $\abs{v_{in}}$, then by setting them equal we again end up with an energy polynomial. For the special case where $m$ is even and $n$ is odd or the other way around, the constant term of the polynomial is a product of the couplings. Since all the eigenvalues have to satisfy the polynomial equation, the zero energy has to do so too. However, this would mean that at least one of the couplings has to be equal to zero and consequently that the chain gets disconnected. In conclusion, for odd-sized chains no PST is possible between even and odd sites.

Things get more involved when we try to rule out other PST's that do not fall into the two cases we have mentioned so far. To this purpose the second reachability criterium has to be employed. We will explicitly demonstrate an analytical scheme that can be used for these cases by considering a specific example. Namely, we will rule out a PST between the first and the fourth site of a 6-site chain. The energy polynomial in this case, when we set $\abs{v_{i4}}$ equal to $\abs{v_{i6}}$ is
\begin{equation}
\label{eq:fx}
E^{3}-(J_{1}^{2}+J_{2}^{2})E+s_{14} J_{1}J_{2}J_{3}=0,
\end{equation} 
where $s_{14}=\pm$. By using Descartes rule we can deduce the maximum number of the polynomial's positive roots depending on the sign of the constant term. In this particular case, since we are dealing with a 6-site chain and two third degree polynomials, all the roots have to be eigenenergies of the system. For $s_{14}=+$ the polynomial can have two real positive roots $E_{1}$, $E_{2}$ and one negative $E_{3}$. While, for $s_{14}=-$, we get one real positive $E_{4}$ and two negative $E_{5}$, $E_{6}$. Due to the symmetry of the spectrum we also get, that $E_{6}=-E_{1}$, $E_{5}=-E_{2}$ and $E_{3}=-E_{4}$. Considering the above, it is straightforward to see that Eq. \ref{eq:amp} becomes a sum of sines.
\begin{equation}
\begin{split}
\bra{4} e^{-iH_{6}t} \ket{1}= & -2 \abs{v_{11}}^{2} \sin{E_{1}t} - 2 \abs{v_{21}}^{2} \sin{E_{2}t} \\
                                                 & +2 \abs{v_{31}}^{2} \sin{E_{3}t}.
\end{split}
\end{equation}
By setting the retrieval time $t^{*}=\pi /2$, due to Eq. \ref{eq:cases}, all the eigenenergies have to be odd integers. On the other hand, employing Vieta's formula, the following equality holds for the three roots of the polynomials.
\begin{equation}
E_{1}=-(E_{2}+E_{3})
\end{equation}
Which in turn implies that $E_{1}$ is an even number as a sum of two odd ones. This proves by contradiction that the two reachability criteria cannot be met simultaneously and PST is not possible for this transfer.

The main point we want to highlight can be stated as follows: when the number of roots of the energy polynomials is greater than the number of the eigenvalues, the second criterium can be used in order to prove that some of these roots cannot satisfy the two reachability criteria simultaneously. It is also clear that as the length of the chain grows the number of these cases is increased, since we are forced to deal with polynomials of greater degree. We have analytically examined open chains up to 10 sites and the optimization algorithm we have used, comes in complete agreement with our analytical findings. Even though the properties of the eigenvalues of Jacobi matrices have been studied extensively ~\cite{friedland,van,da}, the mathematical task to prove that a number of roots of an energy polynomial of arbitrary degree cannot satisfy the rationality criterium, to our knowledge has not yet been properly addressed. As a result, there is no universal analytical procedure that can be followed to deduce whether PST betweeen two states is possible or not. It follows that in general, each case has to be studied separately, which is something that at first sight seems discouraging. Nevertheless, the power of the analytical approach we just presented here is that it can be applied with small modifications for each case and in order to demonstrate this fact more transparently, we have included in Appendix \ref{excl} one more example. 

To sum up, the following general statements can be made for open chains of arbitrary length. PST is always possible between mirror symmetric sites and this has been rigorously proven for the transfer between $1$ and $N$ ~\cite{vinet1}. In addition, transfers where both the initial and target sites are located at the first half of the chain cannot support PST. The same holds for their mirror symmetric counterparts. Finally, for odd-sized chains we have proved that no PST is realized between even and odd sites. To these statements we will also add one that is based on our numerical results. Having examined open chains of length up to $N=20$ sites, we have numerical evidence which support that for even-sized chains PST between the first and $N-1$ site is always possible. The rest of the cases have to be studied separately. If we are unable to prove by contradiction that PST is not supported, we have to search the parametric space and find the suitable profile for the $J$'s that extremizes the fidelity in a finite amount of time.

This can be done numerically via an optimization algorithm or by using yet again the linear system of Eq. (\ref{eq:sysopen}) to analytically extract the optimized profile for the $J_{i}$'s. For the sake of illustration and to gain a more intuitive picture of the physical system under consideration, we will present an indicative example for both cases.

\begin{figure}[h]
\center
\includegraphics[width=0.4\textwidth]{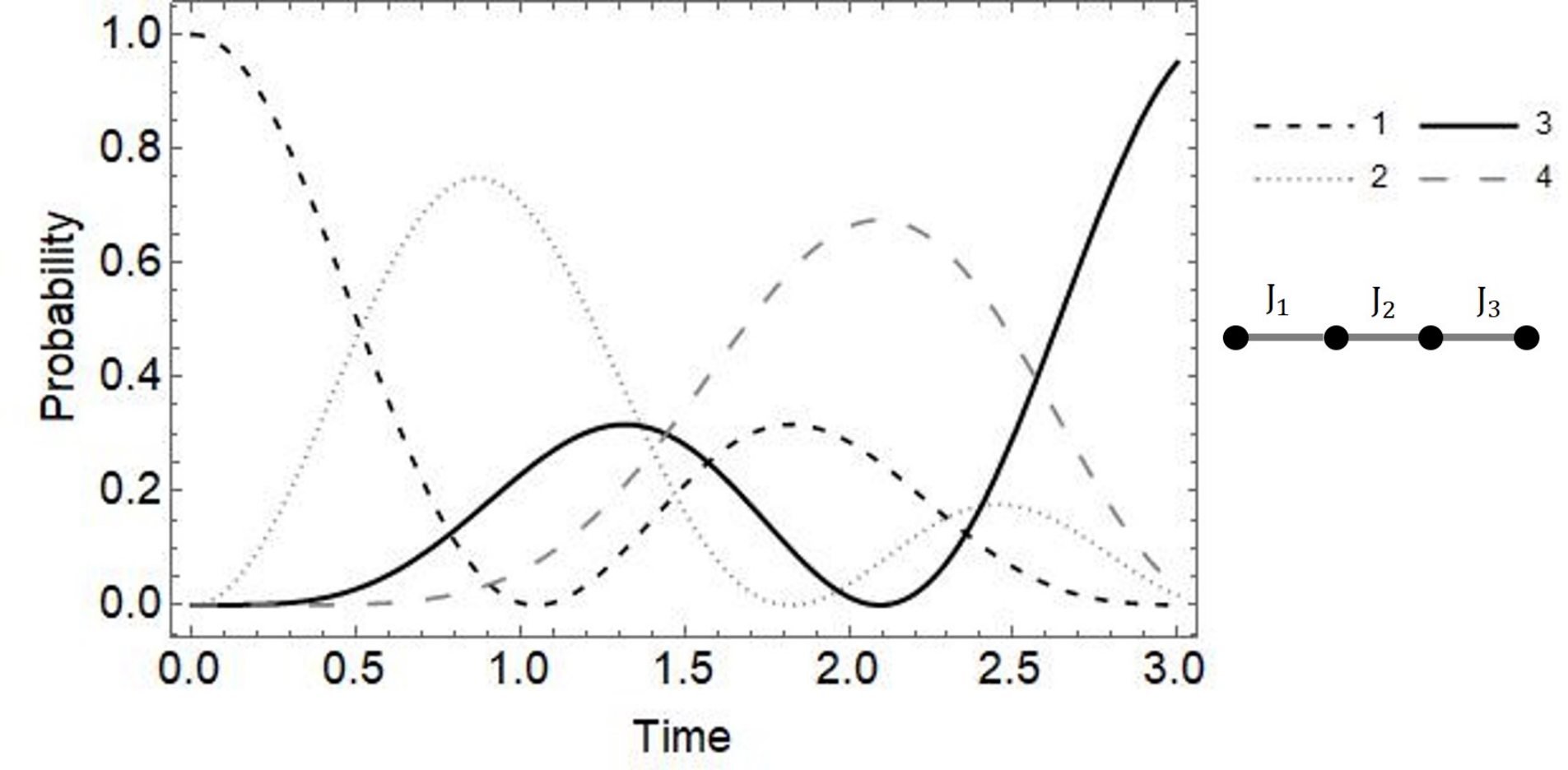}\caption{\label{fig:2} Probability for each lattice site as a function of time for a PST between sites $1$ and $3$. We have set the retrieval time $t^{*}=\pi$ and subsequently the optimal values for the couplings are found to be $J_{1}=1.58114$, $J_{2}=0.948683$ and $J_{3}=1.26491$.}
\centering
\end{figure}

The first example considers the transfer from the first to the third lattice site of an open chain of length $N=4$. This example highlights the fact that the profile of the couplings does not have to be necessarily periodic, as in the case of mirror symmetric lattice sites. In Fig. (\ref{fig:2}) we have plotted the probability for each of the four states on the lattice basis as a function of time, obtained by runninig the optimization algorithm. The system starts at the first site and then gradually the probability spreads out all over the chain, until the whole wavefunction gets localized on the third site at the retrieval time.  
At this point we demonstrate that the method we suggest here to examine reachability, is also very powerfull for designing the optimal profile in reachable cases. By employing the linear system of Eq. (\ref{eq:sysopen}) for a 4-site chain together with the first reachability criterium we obtain the following energy polynomial
\begin{equation}
E^{2}-t_{1}^{2} + s_{13} t_{1}t_{2}=0.
\end{equation}
The system has four eigenenergies that are symmetric around zero, that is $\pm E_{1}$ and $\pm E_{2}$. It is clear that the $s_{13}=+$ gives the pair of eigenenergies with the minimum absolute value, say $\pm E_{1}$, while $s_{13}=-$ corresponds to $\pm E_{2}$. Having in mind the above, it is straightforward to see that
\begin{equation}
\bra{3}e^{-i \tau \mathcal{H}_{N}}\ket{1}=2\abs{v_{11}}^{2} \cos{(E_{1}\tau)}-2\abs{v_{21}}^{2}\cos{(E_{2}\tau)}.
\end{equation}
Setting the retrieval time $\tau=t^{*}=\pi$, forces the eigenenergies to acquire integer values and the first pair that gives an overall sign to the sum is $E_{1}=1$ and $E_{2}=2$. By plugging these energy values into the linear system Eq. (\ref{eq:sysopen}) and solving for the associated coupling values we obtain exactly the same values as those produced running the optimization algorithm.

\subsection{Closed chains}

For closed geometries, the introduction of the coupling between the first and the last site changes the system's behavior in a drastic manner. The linear system in this case takes the following form:
\begin{equation}
\label{eq:sysclosed}
\begin{array}{lcr}
J_{1}s_{i2}\abs{v_{i2}}+J_{N}s_{iN}\abs{v_{iN}}  = E_{i}s_{i1}\abs{v_{i1}} \\
J_{1}s_{i1}\abs{v_{i1}}+J_{2}s_{i3}\abs{v_{i3}}  =  E_{i}s_{i2}\abs{v_{i2}}\\
\vdots \\
J_{N}s{i1}\abs{v_{i1}}+J_{N-1}s_{iN-1}\abs{v_{iN-1}}=E_{i}s_{iN}\abs{v_{iN}}.
\end{array}
\end{equation}
Let us consider PST between an arbitrary pair of sites for a closed chain of fixed length. Following the same procedure as we did for the open chains, we express the eigenvector component of the initial site as a function of the eigenvector component of the target site and we extract two energy polynomials. Due to the cyclic symmetry of the closed system, the degree of the energy polynomials is the same, independently of the choice of the initial and target sites. Namely, the highest degree polynomial for a circular chain of length $N$ is $N-2$ for even-sized chains and $N-1$ for the odd ones. 

From this perspective, it should come as no suprise that our numerical and analytical findings support the following statement: ``For any closed chain of fixed length, if we can find an optimal profile for the couplings that supports PST between a particular pair of sites, then an optimal profile that supports PST between an arbitrary pair of sites always exists". Similarly, if PST is not possible for a pair of sites then the same holds for all pair of sites. Note here that we have assumed different initial and target sites. We do not take into consideration the case of quantum revivals which are always reachable.

In addition to the aforementioned facts, the even or odd length of the chain turns out to play a crucial role to the reachability of a transfer. In particular, all odd-sized chains with the exception of $N=3$ do not support PST. On the contrary, for even-sized chains of arbitrary length, we can always obtain an optimal profile for the couplings that makes the transfer between any particular pair of states reachable.

The $N=3$ closed chain is the only odd geometry in which PST is possible between all pair of sites. We will impose the first reachability criterium on the linear system for the transfer between a pair of sites. Without loss of generality, we pick the first and the third site. Then, depending on which equations we use, we can either obtain a second degree polynomial
\begin{equation}
\label{eq:p1}
E^{2}-s_{13}J_{3}E-J_{1}(J_{1}+s_{13}J_{2})=0,
\end{equation}
or a first degree polynomial
\begin{equation}
\label{eq:p2}
(J_{2}-s_{13}J_{1})E+J_{3}(J_{1}-s_{13})=0.
\end{equation} 
In Eq. (\ref{eq:p1}), we expect that one choice of the sign $s_{13}$ will give one eigenvalue and two more will come from the other. On the other hand, by observing Eq. (\ref{eq:p2}) we could immediately state that, since two first degree polynomials cannot give three solutions, PST is not possible. This however is not the case here. For $s_{13}=-$ we get that $E_{1}=-J_{3}$ but for $s_{13}=+$  we can pick $J_{1}=J_{2}$ which gives an infinite number of solutions and thus we can avoid the contradiction. In conclusion, PST is realized in this system as long as $E_{1}=-J_{3}$ and $J_{1}=J_{2}$. The $N=3$ closed geometry is the only case where a specific choice of the couplings can lead to an omission of the highest order term in the energy polynomial. For all the other odd closed chains ($N>3$), PST is not supported. To analytically demonstrate this fact, we can demand that the energy polynomials, obtained from the linear system (\ref{eq:sysclosed}), possess as roots the system's eigenvalues, arriving this way to a contradiction. For the even-sized closed chains, the optimal profile for the couplings that makes a PST betweeen a particular pair of states reachable, can be obtained by the same scheme that was developed in the previous section.

Nonetheless, the optimization algorithm, as the system's length grows, remains our strongest tool for obtaining the optimal profile for the couplings. Thus, it is worth highlighting a property, that besides its physical importance, enables us to make the optimization algorithm more efficient. When running the algorithm we obtain many solutions for the coupling's profile in reachable cases. Of particular importance is the fact that, there always exists a solution which is locally symmetric on the two different ``paths" (clockwise, anti-clockwise) leading from the initial to the target site. To make this point clear, we will consider a specific example.

\begin{figure}[h]
\center
\includegraphics[width=0.4\textwidth]{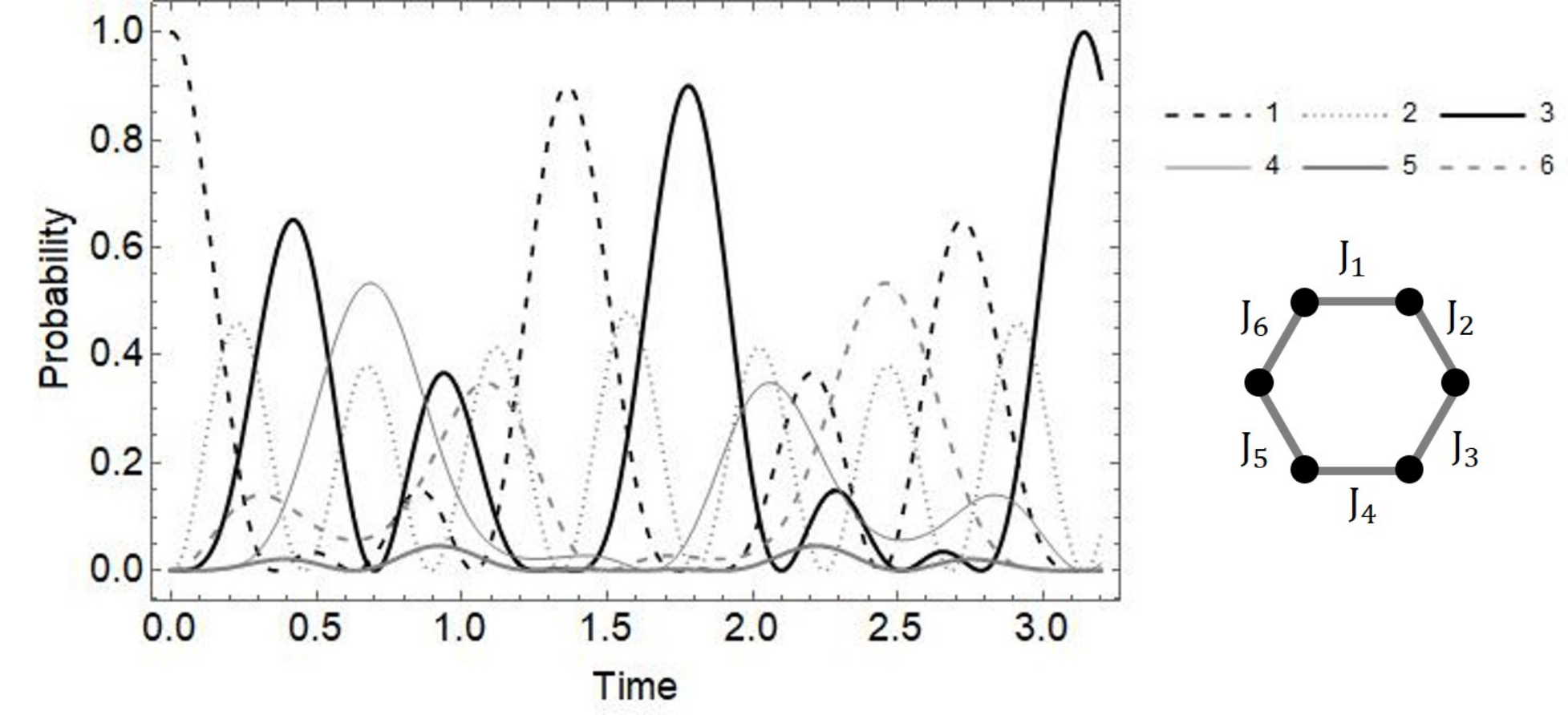}\caption{\label{fig:3} Probability for each lattice site as a function of time for a PST between site $1$ and $3$. We have set the retrieval time $t^{*}=\pi$ and subsequently the optimal values for the couplings are found to be $J_{1}=J_{2}=4.688846$, $J_{3}=J_{6}=2.0$ and $J_{4}=J_{5}=2.239356$.}
\centering
\end{figure}

In Fig. (\ref{fig:3}) we show the probability for each lattice site for an engineered profile of the couplings, that supports PST between the first and the third site of a 6-site closed chain.

By observing the values of the couplings we can easily notice that, for the path that goes clockwise from the first to the third lattice site, $J_{1}=J_{2}$. For the anti-clockwise path, the profile is again parity symmetric ($J_{3}=J_{6}$ and $J_{4}=J_{5}$). Therefore, by imposing such symmetries on the couplings, we can drastically reduce the dimensions of the parametric space in which the optimization algorithm searches for solutions.

\section{Concluding remarks} \label{Con}
In this paper we have derived two reachability criteria that have to be satisfied for perfectly transfering a state between two arbitrary lattice sites of an one dimensional spin-1/2 chain. For open chains of arbitrary length, we have provided a mathematical framework to deduce when PST is possible depending on the size of the chain. For the cases that this was not straightforward, we have developed a scheme that sheds light to the mathematical complexity of the problem which increases with the chain's size. Our results are supported by the numerical implementation of an optimization algorithm from which we can extract the profile of the couplings. By considering closed geometries, we highlighted the ability of even-sized chains to support PST between any pair of sites. This, makes them promising candidates for the realization of an efficient quantum circuit. Our work paves the way towards the completion of a solid mathematical framework for dealing analytically with chains of arbitrary size. In addition, opens the prospect of suitably constructing an optimized profile for the couplings that will enable us to create quantum logic gates for performing operations upon the states.

\section*{Acknowledgments}
N. E. P. gratefully acknowledges financial support from the Hellenic Foundation for Research and Innovation (HFRI) and the General Secretariat for Research and Technology (GSRT), under the HFRI PhD Fellowship grant (GA. no. $74148/2017$). I.B. acknowledges financial support by Greece and the European Union (European Social Fund- ESF) through the Operational Programme ``Human Resources Development, Education and Lifelong Learning" in the context of the project ``Reinforcement of Postdoctoral Researchers” (MIS-$5001552$), implemented by the State Scholarships Foundation (IKY).

\appendix

\section{Proof of the 1st criterium}
\label{proof}

Due to the standard normalization condition it holds:
\begin{equation}
\label{eq:A1}
\sum_{i=1}^{N} \abs{v_{im}}^{2}=1
\end{equation}
Using the Lagrange multipliers method we want to examine under which conditions the quantity $\sum_{i=1}^{N}\abs{v_{im}}\abs{v_{in}}$ is extremized. Thus, we get:
\begin{equation}
\mathcal{Q}=\sum_{i=1}^{N}\abs{v_{im}}\abs{v_{in}}+\lambda_{1}\sum_{i=1}^{N} \abs{v_{im}}^{2} +\lambda_{2}\sum_{i=1}^{N} \abs{v_{in}}^{2}
\end{equation}
Taking the partial derivatives with respect to $\abs{v_{im}}$ and $\abs{v_{in}}$ and setting them to be zero, yields:
\begin{equation}
\label{eq:A3}
\abs{v_{im}}-2\lambda_{1}\abs{v_{in}}=0=\abs{v_{in}}-2\lambda_{2}\abs{v_{im}}
\end{equation}
which results to $4\lambda_{1}\lambda_{2}=1$. Moreover:
\begin{equation}
\sum_{i=1}^{N} \abs{v_{im}}^{2}=4\lambda_{2}^{2}\sum_{i=1}^{N} \abs{v_{in}}^{2}
\end{equation}
which, by taking into account Eq.(\ref{eq:A1}), results to $\lambda_{2}=\pm1/2$. Thus $\lambda_{1}=\lambda_{2}=1/2$, in order for $\mathcal{Q}$ to take its maximum values which means, according to Eq. (\ref{eq:A3}), that $\abs{v_{im}}=\abs{v_{in}}$.

\section{Exclusion scheme example}
\label{excl}

We willl consider the transfer between the first and the fifth site of a $7$-site open chain. Using the linear system of Eq. (\ref{eq:sysopen}), we express $\abs{v_{i5}}$ in terms of $\abs{v_{i1}}$. Setting them equal we obtain two energy polynomials corresponding to $s_{15}=\pm 1$.
\begin{equation}
\label{eq:excl1}
E^{4}-(J_{1}^{2}+J_{2}^{2}+J_{3}^{2})E^{2}+J_{1}^{2}J_{3}^{2}+s_{15} J_{1}J_{2}J_{3}J_{4}=0
\end{equation}
For $s_{15}=+1$ we find four real roots, $\pm E_{1}$ and $\pm E_{2}$. While, for $s_{15}=-1$ a double root $E_{4}=0$ and $E_{3}^{2}=J_{1}^{2}+J_{2}^{2}+J_{3}^{2}$ are obtained. Taking into account the above facts, the probability amplitude of finding the wavefunction localized at the fifth site after time $t$ is given by
\begin{equation}
\begin{split}
\bra{5} e^{-iH_{7}t} \ket{1}= & \abs{v_{11}}^2 e^{-iE_{1}t} + \abs{v_{21}}^{2} e^{-iE_{2}t} -\abs{v_{31}}^{2} e^{-iE_{3}t} \\
    & - \abs{v_{41}}^{2}\\
    & - \abs{v_{51}}^{2} e^{iE_{3}t}- \abs{v_{61}}^{2} e^{iE_{2}t} + \abs{v_{71}} e^{iE_{1}t}.
\end{split}
\end{equation}
 Due to the symmetry of the energy spectrum it also holds that $\abs{v_{11}}=\abs{v_{71}}$, $\abs{v_{21}}=\abs{v_{61}}$ and $\abs{v_{31}}=\abs{v_{51}}$. Thus, it follows that
\begin{equation}
\begin{split}
\bra{5} e^{-iH_{7}t} \ket{1}= & 2 \abs{v_{11}}^{2} \cos{E_{1}t} + 2 \abs{v_{21}}^{2} \cos{E_{2}t} \\
                                                 & - 2 \abs{v_{31}}^{2} \cos{E_{3}t} - \abs{v_{41}}^{2}.
\end{split}
\end{equation}
Without loss of generality we can set $t=\pi$. For the amplitude to get its maximum values, an overall minus sign has to be produced from the cosines. This means that $E_{1}, E_{2}$ have to be odd while $E_{3}$ even. However, from Eq. (\ref{eq:excl1}), if we employ Vieta's formula, we get
\begin{equation}
E_{1}^{2}+E_{2}^{2}=J_{1}^{2}+J_{2}^{2}+J_{3}^{2}=E_{3}^{2}.
\end{equation}
Since, the sum of the squares of two odd integers cannot be the square of an integer, we have proved by contradiction that the transfer under consideration is not reachable.


\begin{thebibliography}{50}

\bibitem{divincenzo}
D. P. DiVincenzo, \textit{``The physical implementation of quantum computation"}, Fortschr. Phys. $\textbf{48}$ (9‐11), 771-783, (2000).  

\bibitem{bose1}
S. Bose, \textit{``Quantum communication through an unmodulated spin chain'}, Phys. Rev. Lett. \textbf{91}, 207901, (2003).  

\bibitem{christandl1}
M. Christandl, N. Datta, A. Ekert and A. J. Landahl,  \textit{``Perfect state transfer in quantum spin networks"}, Phys. Rev. Lett. \textbf{92}, 187902 (2004).

\bibitem{christandl2}
M. Christandl, N. Datta, T. C. Dorlas, A. Ekert, A. Kay and A. J. Landahl, \textit{``Perfect state transfer of arbitrary states in quantum spin networks"}, Phys. Rev. A \textbf{71}, 032312 (2005).

\bibitem{nikolopoulos1}
G. M. Nikolopoulos, D. Petrosyan and P. Lambropoulos, \textit{``Electron wavepacket propagation in a chain of coupled quantum dots"}, J. Phys.: Condens. Matter \textbf{16}, 4991 (2004).

\bibitem{wang}
Y. Wang, F. Shuang, and H. Rabitz, \textit{``All possible coupling schemes in XY spin chains for perfect state transfer"}, Phys. Rev. A \textbf{84}, 012307 (2011).

\bibitem{vinet1}
L. Vinet and A. Zhedanov, \textit{``How to construct spin chains with perfect state transfer"}, Phys. Rev. A \textbf{85}, 012323
(2012).

\bibitem{albanese1}
C. Albanese, M. Christandl, N. Datta and A. Ekert, \textit{``Mirror inversion of quantum states in linear registers"}, Phys. Rev. Lett. \textbf{93}, 230502 (2004).

\bibitem{bellec}
M. Bellec, G. M. Nikolopoulos and S. Tzortzakis, \textit{``Faithful communication Hamiltonian in photonic lattices"}, Opt. Lett. \textbf{37}, 4504 (2012).

\bibitem{perez}
A. Perez-Leija, R. Keil, A. Kay, H. Moya-Cessa, S. Nolte, L.-C. Kwek, B. M. Rodr{\'\i}guez-Lara, A. Szameit and D. N. Christodoulides, \textit{``Coherent quantum transport in photonic lattices"}, Phys. Rev. A \textbf{87}, 012309 (2013).

\bibitem{chapman}
R. J. Chapman, M. Santandrea, Z. Huang, G. Corrielli, A. Crespi, M.-H. Yung, R. Osellame and A. Peruzzo, \textit{``Experimental perfect state transfer of an entangled photonic qubit"}, Nat. Commun. \textbf{7}, 11339 (2016).

\bibitem{bocchieri}
P. Bocchieri and A. Loinger, \textit{``Quantum recurrence theorem"}, Phys. Rev. \textbf{107}, 337 (1957).

\bibitem{friedland}
S. Friedland and A. Melkman, \textit{``On the eigenvalues of non-negative Jacobi matrices"}, Linear Algebra Appl. \textbf{25}, 239 (1979).

\bibitem{van}
P. Van Moerbeke, \textit{``The spectrum of Jacobi matrices"}, Invent. Math. \textbf{37}, 45 (1976).

\bibitem{da}
C. M. da Fonseca, \textit{``On the location of the eigenvalues of Jacobi matrices"}, Appl. Math. Lett. \textbf{19}, 1168 (2006).


\end{thebibliography}
\end{document}